\documentclass[preprint,12pt]{elsarticle}
\journal{International Journal of Plasticity}

\usepackage{graphics,epsfig}
\usepackage{graphicx}
\usepackage{float}
\usepackage{amssymb,amstext,amsmath}
\usepackage{mathtools}
\usepackage{booktabs}
\usepackage{natbib}
\usepackage[utf8]{inputenc}
\usepackage{epstopdf}
\usepackage{subfig}
\usepackage[font=small,labelfont=bf]{caption}

\usepackage{xcolor}
\usepackage{verbatim}

\makeatletter
\def\@maketitle{%
	\newpage
	\begin{center}%
		\let \footnote \thanks
		{\LARGE \@title \par}%
		\vskip 1.2em%
		{\large
			\lineskip .5em%
			\begin{tabular}[t]{c}%
				\@author
			\end{tabular}\par}%
		\vskip 1em%
		{\large \@date}%
	\end{center}%
	\par
	\vskip 1.2em}
\makeatother

\begin{document}
\newcommand{\balpha}{\boldsymbol{\alpha}}
\newcommand{\bbeta}{\boldsymbol{\beta}}
\newcommand{\bgamma}{\boldsymbol{\gamma}}
\newcommand{\bdelta}{\boldsymbol{\delta}}
\newcommand{\bepsilon}{\boldsymbol{\epsilon}}
\newcommand{\bvarepsilon}{\boldsymbol{\varepsilon}}
\newcommand{\bzeta}{\boldsymbol{\zeta}}
\newcommand{\bfoldeta}{\boldsymbol{\eta}}
\newcommand{\btheta}{\boldsymbol{\theta}}
\newcommand{\bvartheta}{\boldsymbol{\vartheta}}
\newcommand{\biota}{\boldsymbol{\iota}}
\newcommand{\bkappa}{\boldsymbol{\kappa}}
\newcommand{\blambda}{\boldsymbol{\lambda}}
\newcommand{\bmu}{\boldsymbol{\mu}}
\newcommand{\bnu}{\boldsymbol{\nu}}
\newcommand{\bxi}{\boldsymbol{\xi}}
\newcommand{\bpi}{\boldsymbol{\pi}}
\newcommand{\bvarpi}{\boldsymbol{\varpi}}
\newcommand{\brho}{\boldsymbol{\rho}}
\newcommand{\bvarrho}{\boldsymbol{\varrho}}
\newcommand{\bsigma}{\boldsymbol{\sigma}}
\newcommand{\bvarsigma}{\boldsymbol{\varsigma}}
\newcommand{\btau}{\boldsymbol{\tau}}
\newcommand{\bupsilon}{\boldsymbol{\upsilon}}
\newcommand{\bphi}{\boldsymbol{\phi}}
\newcommand{\bvarphi}{\boldsymbol{\varphi}}
\newcommand{\bchi}{\boldsymbol{\chi}}
\newcommand{\bpsi}{\boldsymbol{\psi}}
\newcommand{\bomega}{\boldsymbol{\omega}}
\newcommand{\bGamma}{\boldsymbol{\Gamma}}
\newcommand{\bDelta}{\boldsymbol{\Delta}}
\newcommand{\bTheta}{\boldsymbol{\Theta}}
\newcommand{\bLambda}{\boldsymbol{\Lambda}}
\newcommand{\bXi}{\boldsymbol{\Xi}}
\newcommand{\bPi}{\boldsymbol{\Pi}}
\newcommand{\bSigma}{\boldsymbol{\Sigma}}
\newcommand{\bUpsilon}{\boldsymbol{\Upsilon}}
\newcommand{\bPhi}{\boldsymbol{\Phi}}
\newcommand{\bPsi}{\boldsymbol{\Psi}}
\newcommand{\bOmega}{\boldsymbol{\Omega}}
\newcommand{\ldbracket}{[\![}
\newcommand{\rdbracket}{]\!]}
\newcommand{\ldangle}{\langle \!\langle}
\newcommand{\rdangle}{\rangle\!\rangle}
\def\Xint#1{\mathchoice
   {\XXint\displaystyle\textstyle{#1}}%
   {\XXint\textstyle\scriptstyle{#1}}%
   {\XXint\scriptstyle\scriptscriptstyle{#1}}%
   {\XXint\scriptscriptstyle\scriptscriptstyle{#1}}%
   \!\int}
\def\XXint#1#2#3{{\setbox0=\hbox{$#1{#2#3}{\int}$}
     \vcenter{\hbox{$#2#3$}}\kern-.5\wd0}}
\def\ddashint{\Xint=}
\def\dashint{\Xint-}

\begin{frontmatter}
\title{Thermodynamic theory of dislocation/grain boundary interaction} 
\author{Y. Piao, K.C. Le\footnote{Corresponding author. E-mail: chau.le@rub.de}, }
\address{Lehrstuhl f\"ur Mechanik - Materialtheorie, Ruhr-Universit\"at Bochum, D-44780 Bochum, Germany}

\begin{abstract}
The thermodynamic theory of dislocation/grain boundary interaction, including dislocation pile-up against, absorption by, and transfer through the grain boundary, is developed for nonuniform plastic deformations in polycrystals. The case study is carried out on the boundary conditions affecting work hardening of a bicrystal subjected to plane constrained shear for three types of grain boundaries: (i) impermeable hard grain boundary, (ii) grain boundary that allows dislocation transfer without absorption, (iii) grain boundary that absorbs dislocations and allows them to pass later.
\end{abstract}

\end{frontmatter}

\section{Introduction}
The interaction between dislocations and grain boundaries plays an essential role in the mechanical response of polycrystalline materials to loading. Several outcomes of this interaction are possible. At low dislocation density the grain boundary blocks the movement of dislocations and forces them to pile-up against it, with the byproduct of linearly increased kinematic work hardening. As the dislocation density exceeds some critical value, a certain amount of dislocations is absorbed by the grain boundary, causing an increase in the surface dislocation density and the misorientation between adjacent grains. Dislocations can also be emitted at a later stage of straining in the neighboring grain or transferred through the grain boundary. In the presence of grain boundaries, non-uniform plastic deformation occurs, leading to non-redundant (geometrically necessary) dislocations, so the need for continuum models that include the gradient of plastic deformation becomes evident. \citet{Berdichevsky1967} were the first to introduce the curl of plastic deformation (Nye-Bilby-Kr\"oner's tensor) into the continuum dislocation theory (CDT) and associate it with the density of non-redundant dislocations. They also proposed the thermodynamics of dislocations, using this density of non-redundant dislocations as a state variable in the free energy and giving the variational formulation of CDT (see the further developments in \citep{Berdichevsky2006a,LeGuenther2014}). The main drawback of this thermodynamic approach is the absence of configurational entropy and its dual quantity, the effective temperature of dislocations, introduced a few decades later by \citet{Langer2010}. Together with the density of redundant dislocations, this configurational entropy enables ones to build up the meaningful thermodynamics of dislocations satisfying two universal laws for plastic flows discovered in \citep{Le2020,LangerLe2020}. Recently, one of us extended CDT by introducing the density of redundant dislocations as well as the configurational entropy as additional state variables \citep{Le2018}. He has shown that the new theory, called thermodynamic dislocation theory (TDT), can predict both kinematic and isotropic work hardening. This TDT has been applied to model grain boundaries as hard obstacles that block dislocations, leading to dislocation pile-up, kinematic work hardening, and Bauschinger effect  \citep{LeTran2018,GuentherLe2021}. \citet{PiaoLe2020} further applied the theory to the problem of interaction between screw dislocations and permeable grain boundary based on the experiments performed by \cite{kondo2016}. There, an interfacial dissipation potential is proposed and additional boundary conditions for the grain boundary are derived to model screw dislocation/grain boundary interaction.  

Alternative phenomenological models that include the interaction between dislocations and grain boundaries have been proposed in the context of strain gradient plasticity by \citet{Fleck1994,Aifantis1999,Gao1999,Gurtin2003}, and a large number of investigations have been carried out within this approach. \citet{Fredriksson2007} proposed an energetic model according to which the plastic work is stored as interfacial energy that depends on the jump of plastic slip at the grain boundary, while \citet{Fleck2009} argued that material hardening due to the grain boundary is mainly carried by energy dissipation. Note that both models allow for a discontinuity in the plastic slip at the grain boundary. \citet{Gurtin2008} introduced a Burgers' tensor at the grain boundary, analogous to the Nye-Bilby-Kr\"oner's tensor for bulk crystals, to determine both the interfacial energy and dissipation. One merit of this tensorial measure is that it allows the theory to satisfy the slip transfer criteria. \citet{Van2013} proposed a similar theory that takes into account the criteria of slip transfer, but differs from the formulation of \citet{Gurtin2008}. Inter-grain interaction modules from two theories are compared by \citet{Gottschalk2016}. We also mention the references  \citep{Aifantis2005,Voyiadjis2009,Ekh2011,Voyiadjis2014,Peng2015,Pouriayevali2017} which use the same approach. Similar to the CDT developed in \citep{Berdichevsky1967,Berdichevsky2006a, LeGuenther2014}, this standard strain gradient plasticity approach ignores completely configurational entropy, so none of the work done within this approach is consistent with thermodynamics of dislocations proposed in \citep{Langer2010,Le2020,LangerLe2020}.

The aim of this work is to model the interaction of edge dislocations with the grain boundary in polycrystals within the framework of TDT. In contrast to the interaction between screw dislocations and grain boundaries studied in our previous paper \citep{PiaoLe2020}, the processes of dislocation pile-up against, absorption by, and transfer through the grain boundary may occur in this situation. Using the variational formulation together with the proposed surface energy and dissipation potential, we will derive the interface boundary conditions for these processes. The case study is carried out on the boundary conditions affecting work hardening of a bicrystal subjected to plane constrained shear for three types of grain boundaries: (i) impermeable hard grain boundary, (ii) grain boundary that allows dislocation transfer without absorption, (iii) grain boundary that absorbs dislocations and allows them to pass later. Note that, according to various experimental observations reported e.g. in \citep{Rupert2009,Molodov2011}, the grain boundary could be migrated during this dislocation absorption, but this possible migration is ignored in this continuum model. We show especially the effect of dislocation absorption on work hardening. 

The paper is organized as follows. After this brief Introduction, we lay down the kinematics and thermodynamic framework of TDT in Section \ref{Section TDT}. In Section \ref{Section interface condition}, we analyze the effects of dissipation potential and interfacial energy on the work hardening of materials and derive the conditions at the grain boundary for various processes during plastic deformation. In the subsequent Section \ref{Section Application}, the application of the developed theory to the problem of a bicrystal subjected to plane constrained shear is illustrated and its numerical treatment is presented. Detailed numerical simulations are then performed and the results are discussed. We conclude the paper with a brief summary in Section \ref{Section Conclusion}. 

\section{Formulation for bulk crystals} \label{Section TDT}
\subsection{Kinematics}
In the small strain theory, we ignore the differences between the Lagrangian and Eulerian coordinates and assume that the total displacement gradient (distortion) $\nabla \mathbf{u}$ can be decomposed additively into elastic and plastic counterparts 
\begin{equation*} 
\nabla \mathbf{u}=\boldsymbol{\beta}^\text{e}+\boldsymbol{\beta},
\end{equation*}
where $\mathbf{u}(\mathbf{x})$ is a displacement field, $\mathbf{x}=(x_1,x_2,x_3)$ a position vector of a material point of the body, and $\nabla $ the nabla operator. The incompatible plastic distortion of the crystal having $n$ active slip systems takes the form 
\begin{equation*}
\boldsymbol{\beta}(\mathbf{x})= \sum_{\alpha=1}^{n} \beta^{\alpha}(\mathbf{x}) \mathbf{s}^{\alpha} \otimes \mathbf{m}^{\alpha},
\end{equation*}
with $\mathbf{s}^{\alpha}$ and $\mathbf{m}^{\alpha}$ being the pair of unit vectors denoting the slip direction and the normal to the slip planes of the corresponding $\alpha$-th slip system. The plastic slip of the $\alpha$-th slip system, $\beta^{\alpha}(\mathbf{x})$, is assumed to be continuously differentiable function except at the interfaces or grain boundaries. Since $\mathbf{s}^{\alpha}$ and $\mathbf{m}^{\alpha}$ are mutually orthogonal, the trace of the plastic distortion always vanishes, \text{tr}$\boldsymbol{\beta}=0$, which indicates volume-preserving plastic deformation due to the conservative motion of dislocations. The total strain $\boldsymbol{\varepsilon}$ and the plastic strain $\boldsymbol{\varepsilon}^\text{p}$ are the symmetric part of the displacement gradient and the plastic distortion, respectively, so that 
\begin{equation*}
\boldsymbol{\varepsilon}=\frac{1}{2}( \nabla \mathbf{u}+\mathbf{u}\nabla ), \quad \boldsymbol{\varepsilon}^\text{p}=\frac{1}{2}(\boldsymbol{\beta}+\boldsymbol{\beta}^T).
\end{equation*}
Accordingly, the elastic strain equals 
\begin{equation*} 
\boldsymbol{\varepsilon}^\text{e}=\boldsymbol{\varepsilon}-\boldsymbol{\varepsilon}^\text{p}.
\end{equation*}
The net Burger's vector $\mathbf{B}$ of non-redundant dislocations whose lines cross the area $\mathcal{S}$ bounded by the closed curve $\partial \mathcal{S}$ is given by 
\begin{equation*}
\mathbf{B}= \oint_{\partial \mathcal{S}} \boldsymbol{\beta} \cdot \mathrm{d} \mathbf{x}= \int_\mathcal{S} \boldsymbol{\alpha}\cdot \mathbf{n} \mathrm{d}a,
\end{equation*} 
where $\boldsymbol{\alpha}$ is the Nye-Bilby-Kr\"oner's dislocation tensor \citep{Nye1953,Bilby1955,Kroener1955}, 
\begin{equation*}
\boldsymbol{\alpha}= -\boldsymbol{\beta}\times \nabla, \quad (\alpha_{ij}=\epsilon_{jkl}\beta_{il,k}).
\end{equation*}
In particular, if there is only one active slip system such that $\mathbf{s}$ and $\mathbf{m}$ lie in the $(x_1,x_2)$-plane, while the dislocation lines are parallel to the $x_3$-axis, then $\boldsymbol{\alpha}$ under the plane strain has only two nonzero components $\alpha_{i3}=s_i(\partial_s \beta)$, where $\partial_s$ denotes the derivative in the $\mathbf{s}$-direction. Consequently, the scalar density of the non-redundant dislocations is given by
\begin{displaymath}
\rho^\text{g}=\frac{|\mathrm{d}\mathbf{B}|}{b\mathrm{d}a}=\frac{1}{b}|\partial_s \beta|,
\end{displaymath}
with $b$ being the magnitude of Burger's vector.

\subsection{Thermodynamic framework}
Suppose the material body in form of a slab of constant depth $L$ occupies the 3-D domain $\mathcal{V}$, and its boundary $\partial \mathcal{V}$ consists of two nonintersecting surfaces, $\partial _k$ and $\partial _s$, on which the displacement field $\mathbf{u(x)}$ and the traction-free condition are specified, respectively. Assuming that no body force acts on the crystals, the energy functional reads
\begin{equation}\label{Bulk_energy}
I=\int_{\mathcal{V}} \psi(\boldsymbol{\varepsilon}^\text{e}, \rho^\text{g}, \rho^\text{r}, \chi) \mathrm{d}^3x, 
\end{equation}
where $\mathrm{d}^3x=\mathrm{d}x_1\mathrm{d}x_2\mathrm{d}x_3$ denotes the volume element and $\psi$ is the density of the Helmholtz free energy. The dislocation network is assumed to be two-dimensional, such that the dislocation lines are parallel to the $x_3$-axis and the Burgers' vector of edge dislocations lies in the $(x_1,x_2)$-plane. We let $\rho^\text{r}$ denote the density of redundant dislocations whose resultant Burgers' vector vanishes, and $\rho^\text{g}$ the density of non-redundant dislocations (see subsection 2.1). The sum of the two types of dislocation densities is the total dislocation density $\rho=\rho^\text{r}+\rho^\text{g}$. In the spirit of the dislocation mediated plasticity \citep{Langer2010}, which decomposes the thermodynamic system of dislocated crystal into two subsystems and traces it back to statistical aspects of the defects, $\chi$ is the effective temperature characterizing the slow rearrangement of configuration of atoms during the motion of dislocations, while the ordinary temperature $T$ characterizes the fast vibration of atoms of the kinetic-vibrational subsystem. In some cases, such as thermal softening \citep{Langer2016, LeTranLanger2017,LePiao2019a} or adiabatic shear banding \citep{LeTranLanger2018}, the ordinary temperature plays an important role. However, in the present study we set $T$ to be constant and drop it from the list of arguments of the free energy density $\psi$. The state variables $\boldsymbol{\varepsilon}^\text{e}$ and $\rho^\text{g}$ are dependent variables expressed by $\mathbf{u}$ and $\beta$, while $\rho^\text{r}$ and $\chi$ are independent state variables. We decompose $\psi$ into the part due to the elastic strain, $\psi_\text{e}$, and the remaining parts 
\begin{equation*} 
\psi(\boldsymbol{\varepsilon}^\text{e}, \rho^\text{g}, \rho^\text{r}, \chi)=\psi_\text{e}(\boldsymbol{\varepsilon}^\text{e})+\psi_\text{r}(\rho^\text{r})+\psi_\text{m}(\rho^\text{g})+\psi_\text{c}(\chi, \rho),
\end{equation*}
where $\psi_\text{r}$ is the self-energy density of redundant dislocations, $\psi_\text{m}$ the energy density of the non-redundant dislocations, while $\psi_\text{c}$ the configurational heat \citep{Le2018}. Their expressions are given below
\begin{equation*}
\begin{split}
\psi_\text{e}(\boldsymbol{\varepsilon}^\text{e})&=\frac{1}{2} \lambda (\varepsilon^\text{e}_{kk})^2+\mu \varepsilon^\text{e}_{ij}\varepsilon^\text{e}_{ij}, \quad  \psi_\text{r}(\rho^\text{r})=\gamma_\text{D} \rho^\text{r},\\
\psi _\text{m}(\rho^\text{g})&=\gamma_\text{D} \rho^\text{g}_{ss} \ln \Bigl( \frac{1}{1-\frac{\rho^\text{g}}{\rho^\text{g}_{ss}}}\Bigr),\\
\quad \psi_\text{c}(\chi, \rho)&=-\bar{\chi}(-\rho \ln(a^2 \rho)+\rho), 
\end{split}
\end{equation*}
where $\lambda$ and $\mu$ are Lam$\acute{\text{e}}$ constants, and $\gamma_\text{D}$  the dislocation energy per unit length. Since $-\rho \ln(a^2 \rho)+\rho$ is the entropy per unit area of the two-dimensional dislocation network (see the detailed derivation in \citep{Chowdhury2016}), the ``two-dimensional'' temperature $\bar{\chi}=\chi/L$ is introduced to get the configurational heat having the unit of energy density \citep{Le2018}. Note that $a$ is a length scale of the order of atomic spacing. The defect energy $\psi_\text{m}$ describing the interactions of non-redundant dislocations captures the kinematic hardening. The logarithmic term, originated from \citep{Berdichevsky2006b}, ensures that the defect energy increases linearly for small dislocation density and tends to infinity as $\rho^\text{g}$ approaches the saturated density of non-redundant dislocations $\rho^\text{g}_{ss}$. We assume that $\rho^\text{g}_{ss}=k_0 \rho_{ss}$ is a fraction of $\rho_{ss}$ by coefficient $k_0$, where $\rho_{ss}=(1/a^2)e^{-\gamma_\text{D}/\bar{\chi}}$ is the steady-state dislocation density determined by minimizing the free energy of the configurational subsystem \citep{Langer2010}. Note that $k_0$ is characterized by the interaction between dislocations of the same sign, which is similar to the interaction between charges \citep{Groma2003}. We will see in the subsequent Section the role of $\rho^\text{g}_{ss}$ in the contribution of interfacial energy to the work hardening.   

The applied loading can lead to nucleation, propagation and movement of dislocations, which result in plastic deformations in the crystal. In these processes, dislocations always experience resistance, which causes energy dissipation. The dissipation potential in the bulk has the form
\begin{equation} \label{Bulk_dissipation}
D_\text{b}(\dot{\beta}, \dot{\rho}, \dot{\chi})=\tau_\text{i} \dot{\beta} +\frac{1}{2}d_{\rho}\dot{\rho}^2+\frac{1}{2}d_{\chi}\dot{\chi}^2.
\end{equation}
The first term is the plastic power, where $\tau_\text{i}$ is the internal stress, while the last two terms represent the dissipation due to the multiplication of dislocations and the increase of the configurational temperature \citep{Le2018}. \citet{Langer2010} developed the kinetics of dislocation depinning that dominantly controls the plastic slip rate. Averaging this kinetic equation over the representative volume element, we have established in \citep{LeLeTran2020} that 
\begin{equation*}
\dot{\bar{\beta}}=\frac{q(T,\tau_\text{i},\rho^\text{r})}{t_0}=\frac{b}{t_0}\sqrt{\rho}\left[f_\text{P}(T,\tau_\text{i},\rho^\text{r})-f_\text{P}(T,-\tau_\text{i},\rho^\text{r}) \right]
\end{equation*}
where $t_0$ is a microscopic time scale of the order of the inverse Debye frequency, and
\begin{equation*}
f_\text{P}(T, \tau_\text{i}, \rho^\text{r}) = \exp\Bigl[-\frac{T_\text{P}}{T} \exp\Bigl(-\frac{\tau_\text{i}}{\tau_\text{T}(\rho^\text{r})}\Bigr)\Bigr].
\end{equation*}
Note that $T_\text{P}$ is the activation temperature of the potential well of a pinning site and $\tau_\text{T}(\rho^\text{r})=\mu_\text{T} b \sqrt{\rho^\text{r}}$ is the Taylor stress with $\mu_\text{T}$ being proportional to the shear modulus $\mu$. It was shown that this kinetic equation for $\dot{\bar{\beta}}$ can correctly describe the rate-reducing behavior of isotropic work hardening both when the crystal is loaded on one direction and in the load reversal process \citep{PiaoLe2020}. Provided there is only one slip system inclined at an angle $\phi$ to the $x_1$-axis, then, using similar arguments as in \citep{GuentherLe2021}, we lay down one governing equation for $\dot{\tau}_\text{i}$
\begin{equation}\label{InternalStress} 
\dot{\tau}_\text{i}=\mu\frac{q_0}{t_0}\Bigl(\cos 2\phi -\frac{q(T,\tau_\text{i},\rho^\text{r})}{q_0}\Bigr)
\end{equation}
where $q_0/t_0$ denotes the total shear strain rate. This equation plays the role of the kinematic constraint for the internal stress $\tau_\text{i}$. The remaining equations in the bulk can be derived from the following variational principle: the true displacement field $\check{\mathbf{u}}\mathbf{(x)}$, the true plastic slips $\check{\beta}(\mathbf{x})$, the true density of redundant dislocations $\check{\rho}^\text{r}(\mathbf{x})$, and the true configurational temperature $\check{\chi}(\mathbf{x})$ obey the variational equation
\begin{equation} \label{Sedov_derivation}
\delta I +\int_{\mathcal{V}} \left(\frac{\partial D_\text{b}}{\partial \dot{\beta}} \delta\beta +\frac{\partial D_\text{b}}{\partial \dot{\rho}} \delta\rho+\frac{\partial D_\text{b}}{\partial \dot{\chi}} \delta\chi\right) \mathrm{d}^3x=0
\end{equation}
for all variations of admissible fields $\mathbf{u}(\mathbf{x})$, $\beta(\mathbf{x})$, $\rho^\text{r}(\mathbf{x})$ and $\chi(\mathbf{x})$. Focusing on the contribution of the grain boundary to work  hardening, we assume that the whole external boundary $\partial \mathcal{V}$ admit dislocation to reach it. In this case the plastic slip can be varied arbitrarily at $\partial \mathcal{V}$. The variational equation \eqref{Sedov_derivation} allows one to derive the remaining governing equations. From the variation of $I$ with respect to the displacement field $\mathbf{u(x)}$, we obtain the quasi-static equilibrium equation and the boundary condition \citep{LeSeTran2016}
\begin{equation*}
\begin{split}
\sigma_{ij,j}&=0 \quad \text{ in $\mathcal{V}$},\\
 \sigma_{ij}n_j&=0 \quad \text{on $\partial_s$},
 \end{split}
\end{equation*}
with $\boldsymbol{\sigma}=\partial \psi_\text{e}/\partial \boldsymbol{\varepsilon}^\text{e}$ being the Cauchy stress tensor, and $\mathbf{n}$ the outward unit normal vector to the external boundary. In Eq.~\eqref{Sedov_derivation} the vanishing variation with respect to the plastic slip $\beta(\mathbf{x})$ yields the balance of microforces acting on dislocations \citep{Le2018} and the natural boundary condition at the entire external boundary as
\begin{equation*} 
\begin{split}
\tau-\tau_\text{b}-\tau_\text{i}&=0 \quad \text{ in $\mathcal{V}$},\\
\frac{\partial \psi_\text{m}}{\partial \rho^\text{g}}&=\gamma_\text{D} \quad \text{on $\partial \mathcal{V}$},
 \end{split}
\end{equation*}
where $\tau=s_i \sigma_{ij}m_j$ is the resolved shear stress acting on the slip system, while the back stress is
\begin{equation*} 
\tau_\text{b}=-\frac{1}{b}\Bigl(\frac{\partial \psi_\text{m}}{\partial \rho^\text{g}} \text{sign}\beta_{,s}\Bigr)_{,s}=-\frac{1}{b^2}\frac{\partial ^2\psi_\text{m}}{\partial (\rho^\text{g})^2}\beta_{,ss}.
\end{equation*}
Note that $(.)_{,s}=s_i \partial_i(.)$ is the short-hand notation for $\partial _{s}(.)$. 
Following the suggestions by \cite{Le2018} for functions of $d_{\rho}$ and $d_{\chi}$, the remaining equations obtained from Eq.~\eqref{Sedov_derivation} for $\dot{\chi}$ and $\dot{\rho}$ can be cast into the form
\begin{equation}\label{GoverningEq RhoChi} 
\begin{split}
&\dot{\chi}=\mathcal{K}_{\chi} e_\text{D} \tau_\text{i} \frac{q(T,\tau_\text{i},\rho^\text{r})} {t_0}\Bigl(1-\frac{\chi} {\chi_0}\Bigr),  \\
&\dot {\rho}=\mathcal{K}_{\rho} \frac{\tau_\text{i}}{a^2 \, \nu(T, \rho^\text{r}, q_0)^2} \frac{q(T,\tau_\text{i},\rho^\text{r})}{t_0} \Bigl(1-\frac{\rho}{\rho_{ss}(\chi)}\Bigr).
\end{split}
\end{equation}
Here, $\chi_0$ is a constant denoting the steady-state configurational temperature, while the steady-state dislocation density at a given effective temperature is $\rho_{ss}(\chi)=(1/a^2)e^{-\gamma_\text{D}/\bar{\chi}}$. $\mathcal{K}_{\chi}$ is a factor inversely proportional to the effective specific heat and $\mathcal{K}_{\rho}$ is a energy-conversion coefficient \citep{Langer2017}. The function $\nu(T, \rho^\text{r}, q_0)$ is defined as
\begin{equation*}
\nu(T,\rho^\text{r}, q_0)= \ln\Bigl(\frac{T_\text{P}}{T}\Bigr)-\ln\Bigl[\ln\Bigl(\frac{b\sqrt{\rho^\text{r}}}{q_0}\Bigr)\Bigr].
\end{equation*}

\section{Interface boundary conditions}\label{Section interface condition}
As a prototype, we study a simple problem of plane deformation of a bicrystal whose single grain boundary, which is a plane perpendicular to the $y$-axis, lies at $y=y_\text{g}$ (see Fig.~\ref{Plane constrained}). The extension to the case of polycrystals with multiple grain boundaries can be done without complications. Since the tensorial index notation is no longer required, the coordinates $(x_1,x_2,x_3)$ are changed to $(x,y,z)$ for simplicity. For the sake of definiteness, we assume that only edge dislocations move to and interact with the grain boundary. If the plastic slip $\beta$ depends only on $y$ due to the sample geometry and specific boundary conditions, then the non-redundant dislocation density is proportional to $\beta_{,y}$, $\rho^\text{g}=\varkappa | \beta_{,y}|/ b$, where $\varkappa$ depends on the slip system, as will be discussed in Section \ref{Section Application}.
\subsection{Interfacial energy and dissipation}
Dislocations are long-lived and well-defined defects and have various interactions with the grain boundary, e.g., dislocation pile-up, absorption, and transmission. Within the continuum approach, it should be possible to mimic these interactions by the interfacial energy and dissipation that reflect the underlying physics at the nano/submicron scale. The low-angle grain boundary itself can be regarded as an array of dislocations \citep{Burgers1939} whose interfacial energy depends on the surface dislocation density \citep{Read1950,Vitek1987}. It also acts as an additional barrier to the movement of dislocations and can later absorb or release them. Therefore, in addition to bulk dissipation, interfacial dissipation must also be taken into account. We propose the densities of interfacial energy and dissipation potential as follows
\begin{equation}
\psi_\text{s}(\beta \bigr|_{y_\text{g}\pm0})=\psi_\text{s}(|\ldbracket \beta \rdbracket _{\Gamma}|),
\label{Interface energy}
\end{equation}
and
\begin{equation}
D_\text{s}(\dot{\beta} \bigr|_{y_\text{g}\pm0})=\zeta_\text{y} \frac{|\ldangle \dot{\beta}\rdangle_{\Gamma}|}{b}+\frac{1}{2} \zeta_\text{a}\frac{(\ldbracket \dot{\beta}\rdbracket _{\Gamma})^2}{b^2}. \label{Interface dissipation}
\end{equation}
Here, the vertical line followed by $y_\text{g}\pm0$ indicates the limits of the preceding expression as $y$ approaches the grain boundary position $y_\text{g}$ from above and below, while $\ldangle . \rdangle_{\Gamma}$ and $\ldbracket . \rdbracket _{\Gamma}$ denote the mean-value and the jump of a variable at the interface $\Gamma $, respectively. Thus, according to these definitions
\begin{equation*}
\ldangle\dot{\beta}\rdangle_{\Gamma}=\frac{1}{2}(\dot{\beta}\bigr|_{y_\text{g}+0}+\dot{\beta}\bigr|_{y_\text{g}-0}), \quad \ldbracket \dot{\beta}\rdbracket _{\Gamma}=(\dot{\beta}\bigr|_{y_\text{g}+0}-\dot{\beta}\bigr|_{y_\text{g}-0}).
\end{equation*}
Eq.~\eqref{Interface energy} requires that only the surface dislocation density $|\ldbracket \beta \rdbracket _{\Gamma}|/b$, regarded as the state variable, can be the argument of $\psi_\text{s}$. The first term on the right-hand side of Eq.~\eqref{Interface dissipation} is the dissipation by the rate of the mean plastic slip at the interface, where $\zeta_\text{y}$ plays a role of the yield surface tension. The second term in \eqref{Interface dissipation}, quadratic in $\ldbracket \dot{\beta}\rdbracket _{\Gamma}$, describes rate-dependent dissipation due to the acoustic emission of waves generated during the absorption of dislocations by the grain boundary. 

The boundary conditions involving interfacial energy and dissipation can be derived from the following variational equation
\begin{multline}\label{Variational equation I} 
\delta \Bigl(I+\int_{\Gamma}\psi_\text{s}(|\ldbracket \beta \rdbracket _{\Gamma}|)\mathrm{d}a\Bigr) +\int_{\mathcal{V}} \Bigl(\frac{\partial D_\text{b}}{\partial \dot{\beta}} \delta\beta +\frac{\partial D_\text{b}}{\partial \dot{\rho}} \delta\rho+\frac{\partial D_\text{b}}{\partial \dot{\chi}} \delta\chi\Bigr) \mathrm{d}^3x
\\
+ \int_{\Gamma}\Bigl(\frac{\partial D_\text{s}}{\partial \ldbracket \dot{\beta}\rdbracket _{\Gamma}}\delta \ldbracket \beta \rdbracket _{\Gamma}+\frac{\partial D_\text{s}}{\partial \ldangle \dot{\beta}\rdangle_{\Gamma}} \delta\ldangle\beta\rdangle_{\Gamma}\Bigr)\mathrm{d}a=0,
\end{multline}
which extends Eq.~\eqref{Sedov_derivation} to the case involving the grain boundary, where $\mathrm{d}a$ denotes the area element, while $I$ and $D_\text{b}$ remain exactly the same as in \eqref{Bulk_energy} and \eqref{Bulk_dissipation}, respectively. Edge dislocations may either penetrate the grain boundary or be absorbed by it through dissociation of the dislocations. The latter process requires the continuity of displacement field but causes a jump in plastic slip $\ldbracket \beta\rdbracket_{\Gamma}$ at the interface. This  is revealed by the volume integral containing the variation of the non-redundant dislocation density $\delta \rho^\text{g}$ during the integration by parts, permitting the discontinuity to enter into the interface as
\begin{multline} \label{Integraion by parts wrt rhog}
\int_{\mathcal{V}} \Bigl( \frac{\partial \psi_\text{m}}{\partial \rho^\text{g}}+\frac{\partial \psi_{\chi}}{\partial \rho^\text{g}}+\frac{\partial D_\text{b}}{\partial \dot{\rho}^\text{g}}\Bigr) \delta \rho^\text{g} \mathrm{d}^3x
=-\int_{\mathcal{V}} \frac{\varkappa}{b}\Bigl(\frac{\partial \psi_\text{m}}{\partial \rho^\text{g}}-\gamma_\text{D}\Bigr)_{,y}\text{sign}\beta_{,y} \delta\beta  \mathrm{d}^3x \\
+\int_{\partial_s} \frac{\varkappa}{b}\Bigl(\frac{\partial \psi_\text{m}}{\partial \rho^\text{g}}-\gamma_\text{D}\Bigr)\text{sign}\beta_{,y} \delta\beta  \mathrm{d}a 
-\int_{\Gamma} \ldbracket \frac{\varkappa}{b}\Bigl( \frac{\partial \psi_\text{m}}{\partial \rho^\text{g}}-\gamma_\text{D}\Bigr) \text{sign}\beta_{,y} \delta\beta \rdbracket _{\Gamma} \mathrm{d}a.
\end{multline}
The governing equation with respect to $\rho^\text{r}$ from Eq.~\eqref{Sedov_derivation},  given as \citep{Le2018,Le2019}
\begin{equation*}
\gamma_\text{D}+\chi \ln(a^2 \rho)/L+d_{\rho} \dot{\rho}=0,
\end{equation*}
is utilized for $\gamma_\text{D}$ in the derivation of Eq.~\eqref{Integraion by parts wrt rhog}. Expanding the interface integral term on the right hand side of Eq.~\eqref{Integraion by parts wrt rhog} by means of the jump identity \citep{Abeyaratne1990}
\begin{equation*}
\ldbracket p \, q\rdbracket _{\Gamma}=\ldbracket p\rdbracket _{\Gamma}\ldangle q\rdangle_{\Gamma}+ \ldbracket q\rdbracket _{\Gamma}\ldangle p\rdangle_{\Gamma}
\end{equation*}
and substituting it into Eq.~\eqref{Variational equation I}, we transform the latter to
\begin{multline}\label{Variation}
\int_{\Gamma} \Bigl\{ -\ldangle \frac{\varkappa}{b}\Bigl(\frac{\partial \psi_\text{m}}{\partial \rho^\text{g}}-\gamma_\text{D}\Bigr) \text{sign}\beta_{,y}\rdangle_{\Gamma} \delta\ldbracket \beta\rdbracket_{\Gamma}
-\ldbracket \frac{\varkappa}{b}\Bigl(\frac{\partial \psi_\text{m}}{\partial \rho^\text{g}}-\gamma_\text{D}\Bigr) \text{sign}\beta_{,y}\rdbracket_{\Gamma} \delta\ldangle\beta\rdangle_{\Gamma}
\\
+ \frac{\partial \psi_\text{s}}{\partial \ldbracket\beta\rdbracket_{\Gamma}} \delta\ldbracket\beta\rdbracket_{\Gamma} 
+\zeta_\text{y} \frac{1}{b}\text{sign}\ldangle \dot{\beta}\rdangle_{\Gamma}\delta\ldangle\beta\rdangle_{\Gamma}
+\zeta_\text{a} \frac{1}{b^2}\ldbracket \dot{\beta}\rdbracket _{\Gamma} \delta\ldbracket \beta \rdbracket_{\Gamma}\Bigr\} \mathrm{d}a=0.
\end{multline}
To derive consequences from \eqref{Variation}, we need to analyze the variations of plastic slip on both sides of the grain boundary, which may be subject to constraints depending on the accumulated dislocation densities. We therefore consider two different processes.

\textbf{Pile-up process}: When dislocations move toward the grain boundary under the Peach-Koehler force, they first pile up against it. Since the dislocations cannot initially penetrate the grain boundary acting as an obstacle, the plastic slips on both sides of the grain boundary are subject to homogeneous Dirichlet boundary condition which fulfills Eq.~\eqref{Variation} due to $\delta\beta\bigr|_{y_\text{g}-0}=\delta\beta\bigr|_{y_\text{g}+0}=0$. So we have during the pile-up process
\begin{equation}\label{Pileup condition}
\dot{\beta}(y_\text{g}\pm 0, t)=0.
\end{equation}
The time derivative in Eq.~\eqref{Pileup condition} ensures the continuity of plastic slip at the interface with respect to time in the reversal loading \citep{PiaoLe2020}. 

\textbf{Absorption and transmission processes}: As more dislocations of the same sign pile up against the interface, the back stress near the pile-up sites is increased. When the accumulated non-redundant dislocation density and with it the back stress in the vicinity of the grain boundary exceeds a threshold, the latter can no longer block dislocations. When the grain boundary starts to absorb dislocations, the jump in plastic slip as well as the misorientation between two adjacent grains will increase. The plastic slip at the grain boundary as well as its jump $\ldbracket \beta\rdbracket _{\Gamma}$ may change during the process of dislocation absorption, and consequently,
$\delta\ldangle\beta\rdangle_{\Gamma}$ and $\delta\ldbracket \beta \rdbracket_{\Gamma}$ can be chosen independently and arbitrarily. Eq.~\eqref{Variation} then implies
\begin{equation}\label{Yield}
\varkappa \Bigl(\frac{\partial \psi_\text{m}}{\partial\rho^\text{g}}-\gamma_\text{D}\Bigr) \Bigr|_{y_\text{g}+0}+\varkappa \Bigl(\frac{\partial \psi_\text{m}}{\partial\rho^\text{g}}-\gamma_\text{D}\Bigr) \Bigr|_{y_\text{g}-0}=\zeta_\text{y},
\end{equation}
and 
\begin{equation}
\label{Evolution}
\ldbracket \dot{\beta}\rdbracket _{\Gamma}=\frac{b^2}{\zeta_\text{a}}\Bigl( \ldbracket \frac{\varkappa }{2b}\frac{\partial \psi_\text{m}}{\partial\rho^\text{g}}\rdbracket_{\Gamma} -\frac{\partial \psi_\text{s}}{\partial \ldbracket\beta\rdbracket_{\Gamma}} \Bigr),
\end{equation}
provided $\text{sign}\beta_{,y}|_{y_\text{g}\pm 0}=\pm 1$, $\text{sign}\ldangle \dot{\beta}\rdangle_{\Gamma}=1$ and $\text{sign}\ldbracket \dot{\beta}\rdbracket_{\Gamma}=1$.

For simplicity let us assume the symmetry such that $\varkappa _l=\varkappa _u=\varkappa $. Let $\rho_\text{y}$ be the root of the equation $\partial \psi_m/\partial\rho^\text{g}=\zeta_\text{y}/2\varkappa +\gamma_\text{D}$ which exists and is unique due to the monotonicity of function $\partial \psi_m/\partial\rho^\text{g}$. It turns out that this dislocation density plays the role of the first threshold. If the dislocation density on one side of the grain boundary exceeds this threshold value, Eq.~\eqref{Yield} can only be satisfied if 
\begin{equation}\label{Two side condition1}
\begin{cases}
\rho^\text{g}|_{y_\text{g}+0}=\varkappa | \beta_{,y}|/ b|_{y_\text{g}+0}=\rho_\text{y}+\rho_\text{j}, \\
\rho^\text{g}|_{y_\text{g}-0}=\varkappa | \beta_{,y}|/ b|_{y_\text{g}-0}=\rho_\text{y}-\rho_\text{j}.
\end{cases}
\end{equation}
Indeed, taking into account that $\rho_\text{j}$ is much less than $\rho_\text{y}$, we can expand function $\partial \psi_\text{m}/\partial\rho^\text{g}$ into the Taylor series near $\rho_\text{y}$ and neglect terms of order higher than $\rho_\text{j}$. Then it is easy to see that \eqref{Two side condition1} satisfies Eq.~\eqref{Yield}. 

But there is another threshold that does not permit dislocations to reach the grain boundary even if the dislocation density on one side exceeds $\rho_\text{y}$. The reason lies in the second term on the right-hand side of \eqref{Evolution}. Since $\psi_\text{s}$ is a function of $|\ldbracket \beta \rdbracket _{\Gamma}|$, its derivative is equal to
\begin{displaymath}
\frac{\partial \psi_\text{s}}{\partial \ldbracket\beta\rdbracket_{\Gamma}}=\frac{\partial \psi_\text{s}}{\partial |\ldbracket\beta\rdbracket_{\Gamma}|}\text{sign}\ldbracket\beta\rdbracket_{\Gamma}.
\end{displaymath}
Because the largest absolute value of this derivative is achieved at $\ldbracket\beta\rdbracket_{\Gamma}=\vartheta_0$, with $\vartheta_0$ being the initial misalignment, as long as the first term in the bracket on the right-hand side is less than this value, $\ldbracket \beta\rdbracket _{\Gamma}$ cannot increase. 
To simplify \eqref{Evolution} we take into account Eq.~\eqref{Two side condition1} and the smallness of $\rho_\text{j}$. Expanding $\partial \psi_\text{m}/\partial\rho^\text{g}$ into the Taylor series near $\rho_\text{y}$ we transform \eqref{Evolution} to
\begin{equation}\label{Two side condition2}
\ldbracket \dot{\beta}\rdbracket _{\Gamma}=\frac{b^2}{\zeta_\text{a}} \Bigl( \frac{\varkappa }{b}\frac{\partial ^2\psi_\text{m}}{\partial (\rho^\text{g})^2}(\rho_\text{y}) \rho_\text{j} -\frac{\partial \psi_\text{s}}{\partial |\ldbracket\beta\rdbracket_{\Gamma}|}\text{sign}\ldbracket\beta\rdbracket_{\Gamma} \Bigr).
\end{equation}

For the density of interfacial energy we will consider the modified Read-Shockley's interfacial energy \citep{Read1950}, 
\begin{equation}
\label{RS}
\psi_\text{s}=\frac{\gamma_\text{D}}{4\pi (1-\nu)b}\vartheta \ln \frac{e\vartheta_\text{m}}{\vartheta+\delta}, 
\end{equation}
where $\vartheta=|\ldbracket\beta\rdbracket_{\Gamma}|$. A small number $\delta$ is added to the denominator of the logarithm to make the derivative of $\psi_\text{s}$ finite at $\vartheta=0$. Finally, we need an equation for $\rho_\text{y}$ which is similar to Eq.~\eqref{InternalStress} for $\tau_\text{i}$. As such we propose
\begin{equation}
\label{Rho_y}
\rho_\text{y}=\rho_{\text{y}0}+\kappa_\text{a} |\ldbracket\beta\rdbracket_{\Gamma}|.
\end{equation}

\section{Application}\label{Section Application}
\subsection{Plane constrained shear}
\begin{figure}[htp]
	\includegraphics[width=0.6 \textwidth]{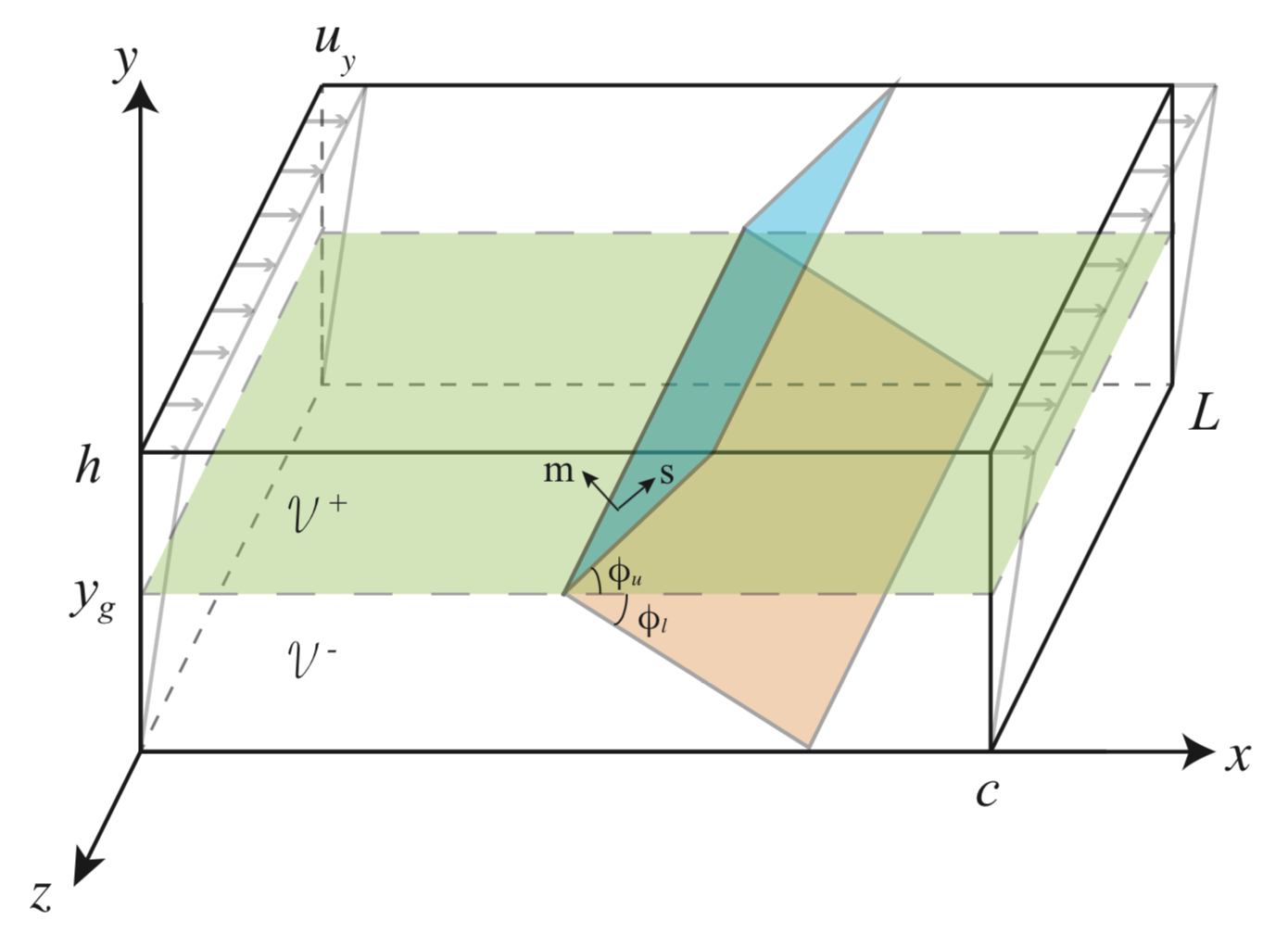} 
	\caption{(Color online) Plane constrained shear of bicrystal}
	\label{Plane constrained}
\end{figure}

Suppose that a bicrystal slab is subjected to plane constrained shear by a given displacement $u_y(t)=\dot{\gamma}t y$ specified at the upper and lower surfaces (see Fig.~\ref{Plane constrained}) and that the system is driven at a constant shear rate $\dot{\gamma }=q_0/t_0$. Let the green colored plane $\Gamma$ be the plane of the grain boundary that is parallel to the $(x,y)$-plane and is located at $y=y_\text{g}$. This grain boundary divides the bicrystal occupying $\mathcal{V}$ into two perfectly bonded single crystals occupying $\mathcal{V}^-$ and $\mathcal{V}^+$, i.e. $\mathcal{V}= \mathcal{V}^- \cup \mathcal{V}^+\cup \Gamma$. Let the cross section of this bicrystal perpendicular to the $z$-axis be a rectangle of width $c$ and height $h$, having the same shape and size over the entire length $L$. We assume $h\ll c \ll L$ to neglect the end effect near $x=0$ and $x=c$. Edge dislocations can occur during the plastic deformation, and we assume that only one slip system is activated in each single crystal, colored blue and red, respectively. The slip directions are perpendicular to the $z$ axis and inclined at angles $\phi_u$ and $\phi_l$ to the plane of the grain boundary.  Given the geometry of the specimen as well as boundary conditions, we assume that the displacements $(u_x,u_y)$ as well as the plastic slip $\beta$ depends only on $y$: $u_x=u_x(y)$, $u_y=u_y(y)$, $\beta=\beta(y)$. The displacement $u_z$ vanishes.

When the plastic deformation occurs, the edge dislocations move on the active slip system with the slip direction $\mathbf{s}=(\cos \phi, \sin \phi, 0)^T$ and the normal to the slip plane $\mathbf{m}=(-\sin\phi, \cos\phi, 0)^T$. The plastic distortion tensor is given by
\begin{equation*}
\boldsymbol{\beta}=\beta(y,t)
\begin{pmatrix}
-\sin\phi\cos\phi& \cos^2\phi &0\\
-\sin^2\phi& \sin\phi\cos\phi &0\\
0&0&0
\end{pmatrix}.
\end{equation*}
The plastic and elastic strain tensor becomes
\begin{align*}
\boldsymbol{\varepsilon}^\text{p}&=\frac{1}{2}\beta(y,t)
\begin{pmatrix}
-\sin2\phi& \cos2\phi &0\\
\cos2\phi& \sin2\phi &0\\
0&0&0
\end{pmatrix}, 
\\ 
\boldsymbol{\varepsilon}^\text{e}&=\frac{1}{2}
\begin{pmatrix}
\beta\sin2\phi& u_{x,y}-\beta\cos2\phi &0\\
u_{x,y}-\beta\cos2\phi& 2u_{y,y}-\beta\sin2\phi &0\\
0&0&0
\end{pmatrix},
\end{align*}
where the displacement field $\mathbf{u}=(u_x, u_y,0)$ is yet unknown. The non-zero components of Nye-Bilby-Kr\"oner's tensor are $\alpha_{xz}=\beta_{,y}\sin\phi \cos\phi$ and $\alpha_{yz}=\beta_{,y}\sin^2\phi$. Therefore the density of the non-redundant dislocation per unit area perpendicular to the $z$-axis equals
\begin{equation*}
\rho^\text{g}=\frac{1}{b}|\boldsymbol{\alpha} \cdot \boldsymbol{e_z }|=\varkappa |\beta_{,y}|/b, \quad \text{where} \quad \varkappa=|\sin\phi |.
\end{equation*}
The total energy functional of a crystal becomes
\begin{equation}\label{Energy functional}
I=cL\int_0^h [\psi_\text{e}(\boldsymbol{\varepsilon}^\text{e})+\gamma_\text{D}\rho^\text{r}+\psi_\text{m}(\rho^\text{g})+\psi_c] \mathrm{d}{y} + cL \psi_\text{i}|_{y=y_\text{g}},
\end{equation}
where $\psi_\text{e}$ is the energy density due to the elastic strain
\begin{equation*}
\psi_\text{e}=\frac{1}{2}\lambda u^2_{y,y}+\frac{1}{2}\mu(u_{x,y}-\beta \cos2\phi)^2+\frac{1}{4}\mu\beta^2\sin^22\phi+\mu\Bigl(u_{y,y}-\frac{1}{2}\beta\sin2\phi\Bigr)^2.
\end{equation*}
Note that the angle $\phi(y)$ is piecewise constant: 
\begin{equation*}
\phi(y)=
\begin{cases}
\phi_\text{l} \quad \text{for } 0<y<y_\text{g}, \\
\phi_\text{u} \quad \text{for } y_\text{g}<y<h.
\end{cases}
\end{equation*}
Taking the variation of functional $I$ with respect to $u_x$ and $u_y$ at fixed $\beta(y)$ we obtain the equilibrium equations which, after integration, lead to
\begin{equation}\label{Variation wrt u}
\begin{split}
u_{x,y}&=\gamma+(\beta-\langle \beta \rangle) \cos2\phi ,\\
u_{y,y}&=\kappa (\beta-\langle \beta \rangle) \sin2\phi ,
\end{split}
\end{equation}
with $\kappa=\frac{\mu}{\lambda+2\mu}$ and $\langle \beta \rangle=\frac{1}{h}\int_0^h \beta \mathrm{d}y$. Inserting Eq.~\eqref{Variation wrt u} into Eq.~\eqref{Energy functional} we reduce the energy functional to 
\begin{multline*}
I=cL\int_0^h\Bigr[\frac{1}{2}\mu \kappa \langle \beta \rangle ^2 \sin^2 2\phi +\frac{1}{2}\mu (\langle \beta \rangle \cos 2\phi -\gamma)^2+\frac{1}{2}\mu (1-\kappa)\beta^2\sin^22\phi +\gamma_\text{D}\rho^\text{r} \\
+\gamma_\text{D} k_0 \rho_{ss} \ln \Bigl(\frac{1}{1-\frac{\rho^\text{g}} {k_0 \rho_{ss}}} \Bigr)-\chi(-\rho \ln(a^2 \rho)+\rho)/L\Bigl]\mathrm{d}y+ cL \psi_\text{i}\Bigl|_{y=y_\text{g}}.
\end{multline*}
Varying this functional with respect to $\beta$ and substituting into \eqref{Variational equation I}, we obtain the equilibrium equation
\begin{equation}\label{Equilibrium Eq}
\tau-\tau_\text{b}-\tau_\text{i}=0,
\end{equation}
with the resolved and back shear stresses being
\begin{equation*} 
\tau=-\mu(\kappa \langle \beta\rangle \sin^2 2\phi +(\langle \beta \rangle \cos 2\phi -\gamma)\cos 2\phi +(1-\kappa)\beta\sin^2 2\phi), 
\end{equation*}
and
\begin{equation*}
\tau_\text{b}=-\frac{C_1}{(1-C_2 |\beta_{,y}|)^2}\beta_{,yy}, \quad C_1=\frac{\gamma_\text{D}}{k_0 \rho_{ss} b^2}\sin^2 \phi, \quad C_2=\frac{1}{k_0 \rho_{ss} b}|\sin \phi|.
\end{equation*}
At the top and bottom surfaces of the sample, the boundary conditions $\partial \psi_m/\partial\rho^\text{g}=\gamma_\text{D}$ must be fulfilled. It is obvious that these imply the Neuman conditions $\beta_{,y}(0,t)=\beta_{,y}(h,t)=0.$ Besides, at $y=y_\text{g}$, the boundary conditions \eqref{Evolution} (or \eqref{Two side condition2}, alternatively), \eqref{Two side condition1}, and \eqref{Rho_y} must be fulfilled. 

\subsection{Rescaled boundary-value problem}
To facilitate numerical integration, we introduce the following rescaled variables and quantities
\begin{gather*}
\tilde{y}=\frac{y}{b}, \quad \tilde{\rho}=a^2 \rho, \quad \tilde{\rho}^\text{r}=a^2 \rho^\text{r}, \quad \tilde{\rho}^\text{g}=b^2\rho^\text{g}=\varkappa |\beta_{,\tilde{y}}|,\quad \tilde{\chi}=\frac{\bar{\chi}}{\gamma_\text{D}}, \quad \theta=\frac{T}{T_\text{P}}, \\
\tilde{\tau}=\frac{\tau}{\mu}, \quad \tilde{\tau}_\text{i}=\frac{\tau_\text{i}}{\mu}, \quad \tilde{\tau}_\text{b}=\frac{\tau_\text{b}}{\mu},
\end{gather*}
where the variable $\tilde{y}$ changes from zero to $\tilde{h}=h/b$. The plastic slip rate is rewritten in the form
\begin{equation*}
\frac{q(T,\tau_\text{i},\rho^\text{r})}{t_0}=\frac{b}{a}\frac{\tilde{q}(\theta, \tilde{\tau}_\text{i}, \tilde{\rho}^\text{r})}{t_0}.
\end{equation*}
in which
\begin{equation*}
\tilde{q}(\theta, \tilde{\tau}_\text{i},\tilde{\rho}^\text{r})=\sqrt{\tilde{\rho}^\text{r}}(\tilde{f}_\text{P}(\theta, \tilde{\tau}_\text{i},\tilde{\rho}^\text{r})-\tilde{f}_\text{P}(\theta, -\tilde{\tau}_\text{i}, \tilde{\rho}^\text{r})).
\end{equation*}
We set $\tilde{\mu}_\text{T}=(b/a)\mu_\text{T}=\mu r$ and assume that $r$ is independent of temperature and strain rate. Then
\begin{equation*}
\tilde{f}_\text{P}(\theta, \tilde{\tau}_\text{i},\tilde{\rho}^\text{r})=\exp \Bigl(-\frac{1}{\theta} \exp\Bigl(-\frac{\tilde{\tau}_\text{i}} {r\sqrt{\tilde{\rho}^\text{r}}}\Bigr)\Bigr).
\end{equation*}
We define $\tilde{q}_0=(a/b)q_0$ such that $q/q_0=\tilde{q}/\tilde{q}_0$, and function $\tilde{\nu}$ becomes
\begin{equation*}
\tilde{\nu}(\theta, \tilde{\rho}^\text{r}, \tilde{q}_0)=\ln\Bigl(\frac{1}{\theta}\Bigr)-\ln\Bigl(\ln\Bigl(\frac{\sqrt{\tilde{\rho}^\text{r}}}{\tilde{q}_0}\Bigr)\Bigr).
\end{equation*}
The dimensionless steady-state quantities are
\begin{equation*}
\tilde{\rho}_{ss}(\tilde{\chi})=\exp\Bigl(-\frac{1}{\tilde{\chi}}\Bigr),\quad \tilde{\chi}_0=\frac{\chi_0}{e_\text{D}}.
\end{equation*}
Using $\tilde{q}$ instead of $q$ as the dimensionless measure of plastic strain rate means that we are effectively rescaling $t_0$ by a factor $b/a$. We assume that $(a/b)t_0=10^{-12}\,$s. 
Since the shear rate $\dot{\gamma}=q_0/t_0$ is constant, we can replace the time derivative by the derivative with respect to $\gamma $ so that $t_0 \partial/\partial t \rightarrow q_0 \partial/\partial \gamma$. Using the introduced dimensionless variables the governing equations in the bulk, Eq.~\eqref{InternalStress}, Eq.~\eqref{GoverningEq RhoChi} and Eq.~\eqref{Equilibrium Eq}, are transformed to the following system of partial differential equations
\begin{equation}\label{System of equation}
\begin{split}
&\tilde{\tau}-\tilde{\tau}_\text{b}-\tilde{\tau}_\text{i}=0, \\
&\frac{\partial \tilde{\tau}_\text{i}}{\partial \gamma}=\cos2\phi-\frac{\tilde{q}(\theta, \tilde{\tau}_\text{i},\tilde{\rho}^\text{r})}{\tilde{q}_0}, \\
&\frac{\partial \tilde{\chi}}{\partial \gamma}=\mathcal{K}_{\chi} \tilde{\tau}_\text{i} \frac{\tilde{q}(\theta,\tilde{\tau}_\text{i},\tilde{\rho}^\text{r})}{\tilde{q}_0}\Bigl(1-\frac{\tilde{\chi}}{\tilde{\chi}_0}\Bigr),\\
&\frac{\partial \tilde{\rho}}{\partial \gamma}=\mathcal{K}_{\rho} \tilde{\tau}_\text{i}\frac{\tilde{q}(\theta, \tilde{\tau}_\text{i},\tilde{\rho}^\text{r})} {\tilde{\nu}(\theta, \tilde{\rho}^\text{r}, \tilde{q}_0)^2\tilde{q}_0} \Bigl(1-\frac{\tilde{\rho}}{\tilde{\rho}_{ss}(\tilde{\chi})}\Bigr), 
\end{split}
\end{equation}
with the rescaled resolved shear stress and back stress being
\begin{gather*}
\tilde{\tau}=-(\kappa \langle \beta\rangle \sin^2 2\phi +(\langle \beta \rangle \cos 2\phi -\gamma)\cos 2\phi +(1-\kappa)\beta\sin^2 2\phi), \\
\tilde{\tau}_b=-\frac{k_1}{(1-k_2 |\beta_{,\tilde{y}}|)^2} \beta_{,\tilde{y}\tilde{y}},
\end{gather*}
where
\begin{equation*}
k_1=\frac{\gamma_\text{D}}{\mu b^2} k \sin^2\phi, \quad k_2=k|\sin\phi|, \quad k=\frac{a^2}{k_0 \tilde{\rho}_{ss}b^2}.
\end{equation*}

During the absorption and transmission process, the density of non-redundant dislocation $\rho^\text{g}$ near the grain boundary hardly reaches the saturated density $\rho^\text{g}_{ss}$. Therefore, using the Taylor expansion, we can approximate the defect energy $\psi_\text{m}$ as follows
\begin{equation*}
\psi_{\text{m}}=\gamma_\text{D}\tilde{\psi}_\text{m}, \quad  \tilde{\psi}_\text{m}=\rho^\text{g}_{ss}\Bigl(\frac{\rho^\text{g}}{\rho^\text{g}_{ss}} +\frac{1}{2}\Bigl(\frac{\rho^\text{g}}{\rho^\text{g}_{ss}}\Bigr)^2\Bigr).
\end{equation*}
By this approximation, the equilibrium equation for microforces \eqref{System of equation}$_1$, a stiff second-order quasi-linear differential equation, can be transformed into an elliptic differential equation
\begin{equation}\label{elliptic equation}
A\beta''+B\beta+C=0,
\end{equation}
which is numerically more stable than the first one when the jump condition is implemented. The prime denotes the derivative of a function with respect to $\tilde{y}$, and 
\begin{equation*}
\begin{split}
A=\frac{\gamma_\text{D}}{\mu b^2} \frac{a^2}{b^2} \frac{1}{k_0 \tilde{\rho}_{ss}} \sin^2\phi, \quad B=(1-\kappa)\sin^2 2\phi, \quad \\
C=-(\kappa\langle\beta\rangle\sin^22\phi+(\langle\beta\rangle\cos 2\phi-\gamma)\cos 2\phi+\tilde{\tau}_{\text{i}}),
\end{split}
\end{equation*}

At the external boundaries Eq.~\eqref{System of equation}$_1$ is subjected to the boundary condition 
\begin{equation}\label{Boundary condition scaled}
	\beta_{,\tilde{y}}(0,\tilde{f})=0, \quad \beta_{,\tilde{y}}(\tilde{h},\tilde{f})=0.
\end{equation}
From \eqref{Evolution} we get the rescaled interface condition at the grain boundary as
\begin{equation}\label{pqGamma}
	\begin{split}
	&\frac{\partial \ldbracket \beta \rdbracket_{\Gamma}}{\partial \gamma}=\frac{1}{\tilde{\zeta}_\text{a}} \left(\frac{\varkappa}{2} \ldbracket \frac{\partial \tilde{\psi}_{\text{m}}}{\partial \tilde{\rho}^{\text{g}}} \rdbracket_{\Gamma} - \frac{\text{sign}\ldbracket \beta \rdbracket_{\Gamma}}{4\pi (1-\nu_0)}\log\frac{\vartheta_{\text{m}}}{|\ldbracket \beta \rdbracket_{\Gamma}|+\delta}\right) , \\ 
	&\ldbracket\beta_{,\tilde{y}}\rdbracket_{\Gamma}=\frac{2}{\varkappa}\tilde{\rho}_{\text{y}}.
	\end{split}
\end{equation} 
where $\tilde{\zeta}_\text{a}=\zeta_\text{a}t_0/(q_0b\,\gamma_D)$ and $\tilde{\rho}_{\text{y}}=b^2\rho_{\text{y}}$. The dimensionless form of Eq.~\eqref{Rho_y} reads
\begin{equation*}
\tilde{\rho}_\text{y}=\tilde{\rho}_{\text{y}0}+\tilde{\kappa}_\text{a} |\ldbracket\beta\rdbracket_{\Gamma}|,
\end{equation*}
where $\tilde{\kappa}_\text{a}=b^2\kappa_\text{a}$.

We also need to pose the initial conditions for $\beta$, $\tilde{\tau}_i$, $\tilde{\rho}$ and $\tilde{\chi}$. For the internal shear stress, we assume that the specimen is not plastically pre-deformed, i.e. $\tilde{\tau}_\text{i}(\tilde{y},0)=0$. As for the plastic slip, two scenarios can be studied: (i) vanishing initial misalignment, (ii) nonzero initial misalignment. The numerical simulations will be performed for the first case where $\beta(\tilde{y},0)=0$. The rescaled dislocation density $\tilde{\rho}(\tilde{y},0)$ and the rescaled disorder  temperature $\tilde{\chi}(\tilde{y},0)$ are assigned reasonable initial values based on various numerical simulations. It should be noted that, from the physical point of view, these initial values depend on the sample preparation. 

\subsection{Numerical implementation}
We discretize the boundary value problem formulated above by a finite difference method called the line method. The latter replaces the spatial derivatives by finite differences, which allow us to transform partial differential equations into a system of ordinary differential equations with respect to $\gamma$. We illustrate this method for the case $\tilde{y}_\text{g}=\tilde{h}/2$. For the case $\tilde{y}_\text{g}\ne \tilde{h}/2$ the discretization can be done in a similar way. Let the interval $0< \tilde{y}< \tilde{h}$ be divided into $2n$ equal subintervals. Then the step in $\tilde{y}$ along the grid becomes $\Delta \tilde{y}=\tilde{h}/2n$. Because of the jump in $\beta $ at the interface, we distinguish the left and right nodes located at the same point $\tilde{y}_\text{g}=n \Delta \tilde{y}$. The coordinates of the nodes on the left interval $0< \tilde{y}< \tilde{h}/2$ are
\begin{displaymath}
\tilde{y}_i=i \Delta \tilde{y},\quad i=0,\ldots,n,
\end{displaymath}
while those on the right interval $\tilde{h}/2< \tilde{y}< \tilde{h}$ are
\begin{displaymath}
\tilde{y}_i=(i-1)\Delta \tilde{y},\quad i=n+1,\ldots, 2n+1.
\end{displaymath}
Thus, altogether we have $2(n+1)$ nodes. The spatial derivatives of the plastic slip, $\beta_{,\tilde{y}}$ and $\beta_{,\tilde{y}\tilde{y}}$, for all internal nodes not lying at the external boundaries or the interface, are approximated as
\begin{equation*}
\frac{\partial \beta(\tilde{y}_i,\gamma)}{\partial \tilde{y}}=\frac{\beta_{i+1}-\beta_{i-1}}{2\Delta \tilde{y}}, \quad 
\frac{\partial^2 \beta(\tilde{y}_i,\gamma)}{\partial \tilde{y}^2}=\frac{\beta_{i+1}-2\beta_i+\beta_{i-1}}{\Delta \tilde{y}^2},
\end{equation*}
where $\beta_i=\beta(\tilde{y}_i,\gamma)$. We decompose the integral $\langle \beta \rangle$ in the resolved shear stress into two integrals that are calculated using the trapezoidal rule. Thus
\begin{equation*}
\langle \beta \rangle=\frac{1}{2n}\Bigl( \frac{1}{2}\beta_0+\beta_1+\ldots+\frac{1}{2}\beta_n+\frac{1}{2}\beta_{n+1}+\beta_{n+2}+\ldots+\frac{1}{2}\beta_{2n+1}\Bigr).
\end{equation*}
After these replacements Eq.~\eqref{System of equation}$_1$ becomes an ordinary differential equation in all internal nodes.

The first derivatives at the external nodes are replaced by
\begin{displaymath}
\frac{\partial \beta(\tilde{y}_0,\gamma)}{\partial \tilde{y}}=\frac{\beta_{1}-\beta_{0}}{\Delta \tilde{y}}, \quad \frac{\partial \beta(\tilde{y}_{2n+1},\gamma)}{\partial \tilde{y}}=\frac{\beta_{2n+1}-\beta_{2n}}{\Delta \tilde{y}}.
\end{displaymath}
Thus, the boundary conditions \eqref{Boundary condition scaled} become
\begin{equation*}
\beta_{0}=\beta_{1}, \quad \beta_{2n+1}=\beta_{2n}.
\end{equation*}
The first left and right derivatives at the interface are replaced by
\begin{displaymath}
\frac{\partial \beta(\tilde{y}_n,\gamma)}{\partial \tilde{y}}=\frac{\beta_{n}-\beta_{n-1}}{\Delta \tilde{y}}, \quad \frac{\partial \beta(\tilde{y}_{n+1},\gamma)}{\partial \tilde{y}}=\frac{\beta_{n+2}-\beta_{n+1}}{\Delta \tilde{y}}.
\end{displaymath}
while the jump in $\beta $ is
\begin{displaymath}
\ldbracket \beta \rdbracket_{\Gamma}=\beta_{n+1}-\beta_n.
\end{displaymath}
These transform Eqs.~\eqref{pqGamma} into two differential-algebraic equations. 

Since the three remaining equations of \eqref{System of equation} do not contain derivatives, they are satisfied node by node. Besides, as the unknown functions are continuous at the interface, we have
\begin{displaymath}
\tilde{\tau}_{\text{i}n}=\tilde{\tau}_{\text{i}n+1},\quad \tilde{\rho}_{n}=\tilde{\rho}_{n+1},\quad \tilde{\chi}_{n}=\tilde{\chi}_{n+1}.
\end{displaymath}
It is easy to see that the system of partial differential equations and boundary conditions becomes a system of $4(2n+1)$ ordinary differential-algebraic equations (DAE) for $4(2n+1)$ unknowns, so the problem is well-posed. For the numerical discretization we choose $n=200$, while for the integration of DAEs with respect to the time-like variable $\gamma$ a step size $\Delta\gamma=10^{-4}$  is chosen. The DAE-system is then integrated by the Matlab integrator \textit{ode15s}. 

\subsection{Numerical results}
In the numerical simulations, we keep the temperature and shear rate constant at $T=298\,$K and $\tilde{q}_0=10^{-13}$. The parameters for the thermodynamic dislocation theory for copper are chosen as presented in \citep{Langer2010}:
\begin{equation*}
	r=0.0323, \quad \theta=0.0073, \quad \mathcal{K}_{\chi}=350,  \quad \mathcal{K}_{\rho}=96.1, \quad \tilde{\chi}_0=0.25.
\end{equation*}
We choose the following parameters for the copper bicrystal
\begin{equation*}
	h=5.1 \, \mu \text{m}, \, b=0.255\, \text{nm}, \, a=10b, \, \mu=50 \, \text{GPa}, \, \nu_0=0.34, \, \phi=30^{\circ}.
\end{equation*}
The active slip systems of two grains in the bicrystal are constrained to symmetry. The initial conditions for dislocation density and configuration temperature are $\tilde{\rho}(0)=6.25 \times 10^{-5}$, $\tilde{\chi}(0)=0.23$ and we assume $\gamma_\text{D}=\mu b^2$. The critical density is set as $\tilde{\rho}_{\text{y}0}= 1.3 \times 10^{-5}$ and the other two interfacial dissipation parameters are given as $\tilde{\zeta}_\text{a}=1.5$ and $\tilde{\kappa}_\text{a}=5 \times 10^{-4}$. The parameters for the defect energy and interfacial energy are chosen as follows: $k_0=0.2$, $\vartheta_{\text{m}}=0.192$, $\delta=1\times10^{-7}$. Unless some parameters are changed for the parameter study, they are kept as default values. 

\begin{figure}[htp]
	\centering
	\begin{tabular}{cc}
		\subfloat[]{\includegraphics[width=0.48 \textwidth]{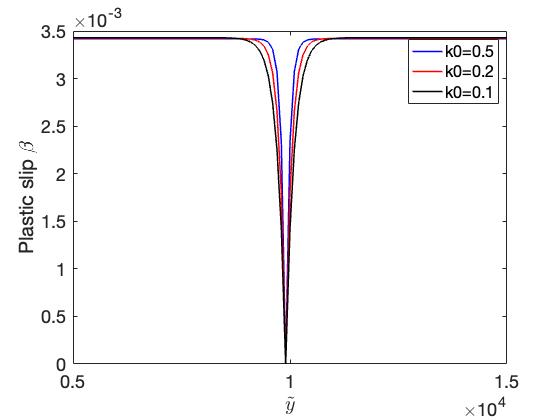}}  &
		\subfloat[]{\includegraphics[width=0.48 \textwidth]{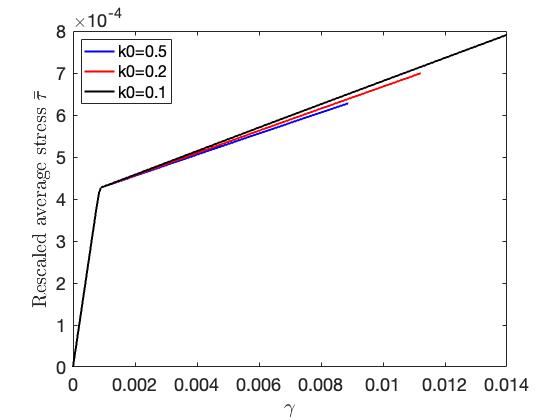}} \\
	\end{tabular}
	\caption{(Color online) (a) Distribution of plastic slips at $\gamma=8*10^{-3}$ for $k_0=0.1$, $k_0=0.2$, and $k_0=0.5$. (b) The corresponding hardening behaviors in the pile-up process.}
	\label{k0_dependence}
\end{figure}

The saturation density of the non-redundant dislocations $\rho^\text{g}_{ss}$ as well as the defect energy $\psi_{\text{m}}$ is characterized by the factor $k_0$, so that its value affects the distribution of plastic slip and work hardening. Fig.~\ref{k0_dependence}(a) shows the distributions of plastic slip in the pile-up process for three different values of $k_0$. When $k_0$ is small, the plastic slip decreases gradually near the grain boundary, while it decreases more steeply to zero for a higher value of $k_0$. Since the slope of plastic slip is proportional to the accumulated dislocation density, these distributions imply that a smaller $k_0$ leads to a larger dislocation accumulation zone and a lower accumulated density of non-redundant dislocations near the grain boundary. This is because a small $k_0$ makes $\rho^{\text{g}}_{ss}$ small, so a lower density of non-redundant dislocations is required for the backstress to have an effect on the balance of micro-forces. The hardening behavior in the pile-up process is shown in Fig.~\ref{k0_dependence}(b). It can be seen that the hardening rate is larger for smaller $k_0$. The reason is that the wider dislocation accumulation zone and the slower gradient increase of plastic slip resulting from the small $k_0$ lead to a higher resolved shear stress and consequently a larger hardening rate. A smaller value of $k_0$ also leads to a longer length of the accumulation phase. This is because the accumulated $\rho^{\text{g}}$ is relatively low for small $k_0$ and therefore it takes longer to reach the required critical density $\rho_{\text{y}}$. 

\begin{figure}[htp]
	\centering
	\subfloat[]{\includegraphics[width=0.48 \textwidth]{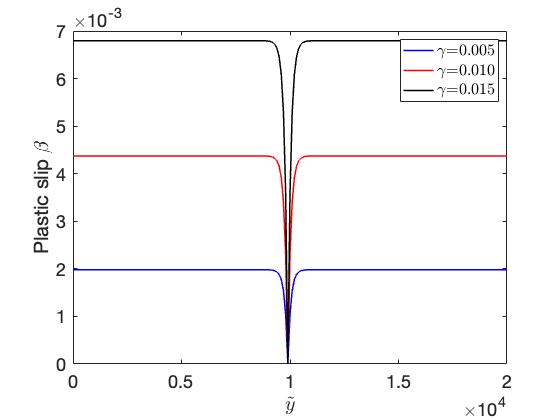}}
	\subfloat[]{\includegraphics[width=0.48\textwidth]{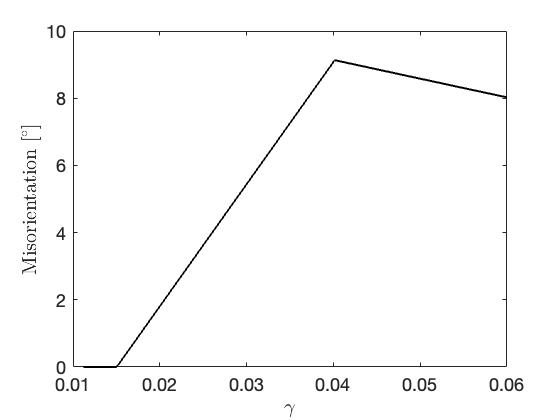}}
	\caption{(Color online) (a) Plastic slips during the pile-up process, (b) Evolution of the misorientation at the grain boundary.}
	\label{Beta_JumpEvolution}
\end{figure}

At different shear strains $\gamma$, the distribution of plastic slip in the processes of dislocation pile-up is shown in Fig.~\ref{Beta_JumpEvolution}(a). The Neumann boundary conditions $\beta_{,\tilde{y}}(0)= \beta_{,\tilde{y}}(\tilde{h})=0$ cause the curve of plastic slip to remain flat on two sides, while a groove is formed at the position of the grain boundary due to the Dirichlet interface condition. A dislocation pile-up zone is located in the center and two dislocation-free zones are located at the boundary layers. In this process, the homogeneous Dirichlet interface condition is applied, and the dislocations cannot penetrate into the grain boundary. Therefore, the groove sinks with increasing strain, but the tip remains attached to the bottom (Fig.~\ref{Beta_JumpEvolution}(a)). Fig.~\ref{Beta_JumpEvolution}(b) shows the evolution of the misalignment at the grain boundary, which is obtained from Eq.~\eqref{Evolution}. When the non-redundant dislocation density $\rho^g$ reaches the critical value $\rho_{\text{y}0}$ at $\gamma=0.011$, the evolution equation is activated. The grain boundary starts absorbing dislocations and the misorientation increases only when the strain exceeds the second threshold value at $\gamma=0.0151$, up to which the accumulation process continues. The dislocation absorption process stops at $\gamma=0.0403$ when the slope of the misorientation changes. 

\begin{figure}[htp]
	\centering
	\subfloat[]{\includegraphics[width=0.48 \textwidth]{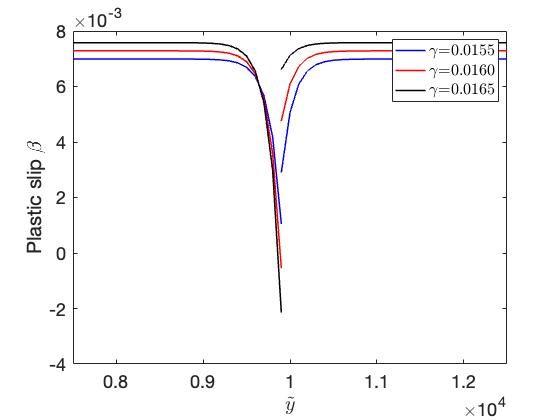}}
	\subfloat[]{\includegraphics[width=0.48\textwidth]{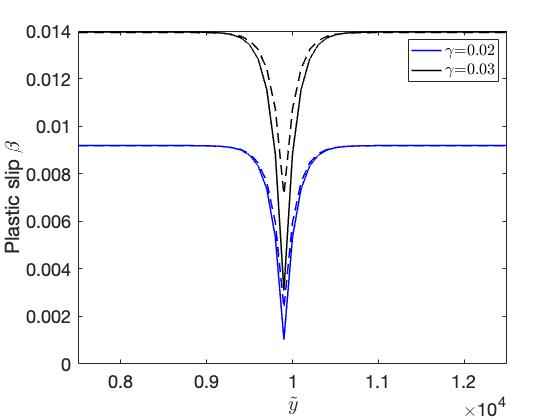}}
	\caption{(Color online) (a) Plastic slips for absorption process, (b) Plastic slip omitting the jump.}
	\label{PlasticSlip_Absorption}
\end{figure}

The distribution of plastic slip during the dislocation absorption process is shown in Fig.~\ref{PlasticSlip_Absorption}(a). In this process, the grain boundary is no longer impenetrable but absorbs dislocations, increasing the misorientation and causing non-redundant dislocations to accumulate. These are two factors that cause the distribution of plastic slip to take a different form than in the pile-up process, as shown in Fig.~\ref{PlasticSlip_Absorption}(a). Here, it can be observed that the plastic slip on two sides of the grain boundary converges not to a single point, but to two points apart, and the distance between them increases, which means that the misorientation of the grain boundary increases. To see the effect of the additionally accumulated non-redundant dislocations, we omit the jump from the plastic slip distribution at $\gamma=0.02$ and $\gamma=0.03$ (solid lines) and compare it with that at the beginning of the absorption process (dashed lines) in Fig.~\ref{PlasticSlip_Absorption}(b). It can be seen that the tip leaves the bottom, and as the shear stress increases, the discrepancy between the solid and dashed lines also increases, indicating that the groove does not stop sinking.

\begin{figure}[t]
	\centering
	\subfloat[]{\includegraphics[width=0.48\textwidth]{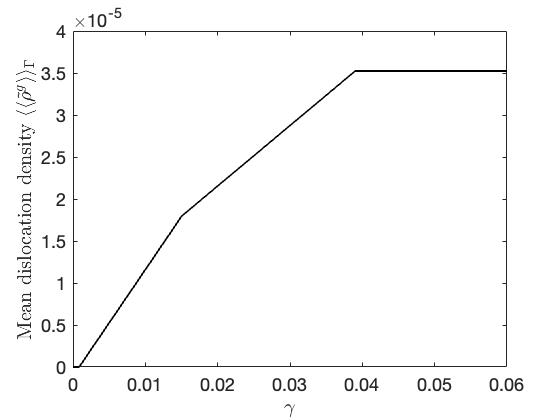}}
	\subfloat[]{\includegraphics[width=0.48\textwidth]{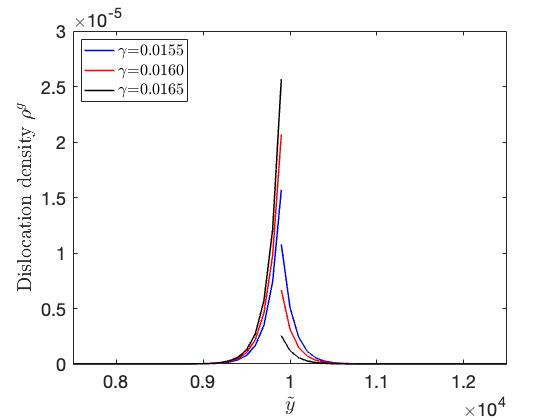}}
	\caption{(Color online) (a) Evolution of the mean density of the non-redundant dislocations, (b) Evolution of the non-redundant dislocation density at two sides of grain boundary.}
	\label{Density_process}
\end{figure}

In the pile-up process, the rate of accumulated dislocation density is the highest among the different processes because all dislocations are blocked by the grain boundary. When positive and negative dislocations enter the grain boundary and increase the misorientation in the absorption process, the ability of the grain boundary to block dislocations from transfer is enhanced, and no additional dislocations are accumulated after this process. This behavior can be seen in Fig.~\ref{Density_process}(a). In the pile-up process, the positive and negative non-redundant dislocation densities are symmetrically distributed on both sides of the grain boundary, while in the absorption process, a jump in the non-redundant density occurs near the grain boundary, as shown in Fig.~\ref{Density_process}(b). Under the interfacial condition Eq.~\eqref{Two side condition1}, the same amount of dislocation density is shifted from one side to the other.

\begin{figure}[htp]
	\centering
	\begin{tabular}{cc}
		\subfloat[]{\includegraphics[width=0.48 \textwidth]{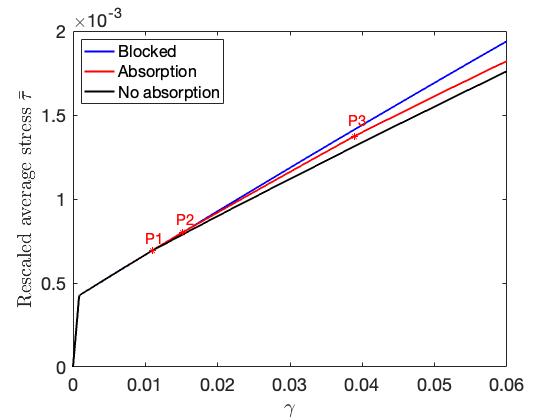}}  &
		\subfloat[]{\includegraphics[width=0.48 \textwidth]{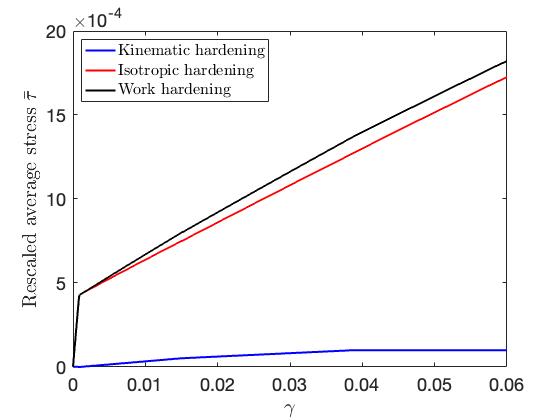}} \\
	\end{tabular}
	\caption{(Color online) Rescaled stress-strain curves: (a) with impenetrable grain boundary (blue), with dislocation absorption process included (red), without dislocation absorption process (black), (b) Partition of work hardening.}
	\label{Hardening_compare}
\end{figure}

In Fig.~\ref{Hardening_compare}(a), the stress-strain curve of the present model (red curve) is shown, where P1 and P2 indicate the first and second thresholds and P3 the end of the absorption process. For comparison we plot also the stress-strain curves obtained by two other models. The blue curve represents the hardening behavior of bicrystals with impenetrable grain boundary, which blocks all non-redundant dislocations and therefore hardens the most. The black curve describes the hardening behavior of bicrystals without absorption process. When $\rho^{\text{g}}$ reaches the critical density $\rho_{\text{y}0}$, the pile-up process ends at P1 and the traversal process follows. These two curves represent the limits above and below which the hardening with absorption can vary due to different interfacial dissipation. Fig.~\ref{Hardening_compare}(b) shows the partition of work hardening into isotropic and kinematic hardening. Isotropic work hardening results from the internal stress required by the dislocations to overcome the resistance due to pinning. Kinematic work hardening is caused by back stresses describing the interaction between non-redundant dislocations. In the non-uniform plastic deformation of small size crystals, kinematic work hardening tends to contribute strongly, while in the present study the contribution is small because bicrystals have only one grain boundary where non-redundant dislocations accumulate. It is worth mentioning that for crystals with multiple grain boundaries, the kinematic work hardening rate is higher \citep{PiaoLe2020}.

\begin{figure}[htp]
	\centering
	\begin{tabular}{cc}
		\subfloat[]{\includegraphics[width=0.48 \textwidth]{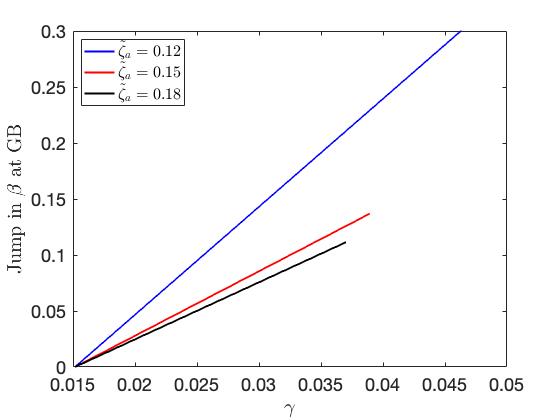}}  &
		\subfloat[]{\includegraphics[width=0.48 \textwidth]{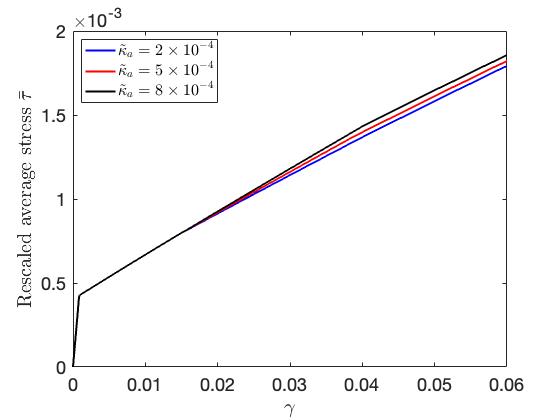}} \\
	\end{tabular}
	\caption{(Color online) (a) Influence of $\tilde{\zeta}_\text{a}$ to the evolution of jump in plastic slip, (b) Influence of $\tilde{\kappa}_\text{a}$ to the hardening behavior.}
	\label{Dissi_Influence}
\end{figure}

Among the interfacial dissipation parameters, $\rho_{\text{y}0}$ determines the first threshold and $\tilde{\zeta}_\text{a}$ the second, so they determine the onset of the dislocation absorption process. The size of the jump in the plastic slip as well as the length of the absorption process are controlled by $\tilde{\zeta}_\text{a}$, as shown in Fig.~\ref{Dissi_Influence}(a). Note that the jump exhibits a linear dependence on the shear strain. The kinematic strain hardening in the absorption process (Fig.~\ref{Dissi_Influence}(b)) is influenced by the additional accumulated non-redundant dislocations originating from the increased misorientation, so that $\tilde{\kappa}_\text{a}$ is the key controlling parameter.

\begin{figure}[htp]
	\centering
	\begin{tabular}{cc}
		\subfloat[]{\includegraphics[width=0.48 \textwidth]{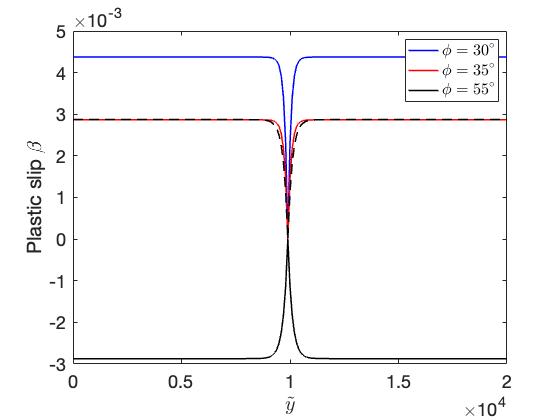}}  &
		\subfloat[]{\includegraphics[width=0.48 \textwidth]{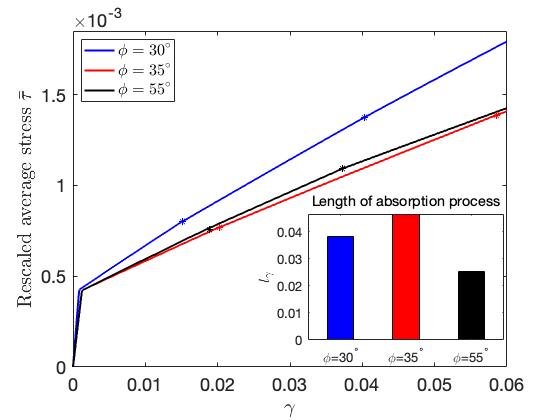}} \\
	\end{tabular}
	\caption{(Color online) (a) Distribution of plastic slips in pile-up process for $\phi=30^{\circ}$, $\phi=35^{\circ}$ and  $\phi=55^{\circ}$, (b) The corresponding hardening behaviors.}
	\label{Hardening_angle}
\end{figure}

Finally, we show the distribution of plastic slip at $\gamma=0.01$ with $\kappa_a=2\times 10^{-4}$ for $\phi=30^{\circ}$, $\phi=35^{\circ}$ and $\phi=55^{\circ}$ in Fig.~\ref{Hardening_angle}(a). It can be seen that for $0<\phi<45^{\circ}$ and $135^{\circ}<\phi<180^{\circ}$ the plastic slip is positively distributed, while for $45^{\circ}<\phi<135^{\circ}$ it is the reverse. 
The dashed curve is symmetrical to the black curve about the horizontal coordinate axis. We see that the blue and dashed curves coincide in the dislocation-free zone. This is because the resolved shear stress $\tau$ and the residual stress $\tau_{\text{i}}$ for $\phi=35^{\circ}$ are identical in magnitude but opposite in sign to those for $\phi=55^{\circ}$. The distributions of plastic slip are different near the grain boundary because the accumulation of non-redundant dislocation density is inconsistent due to the different slip systems. Fig.~\ref{Hardening_angle}(b) shows the strain hardening behavior for three different $\phi$, where the asterisks indicate the beginning and end of the absorption processes. It can be seen that the slip system affects not only the hardening rate but also the length of the absorption process $l_{\gamma}$, as indicated in the bar graph. 

\section{Conclusion}\label{Section Conclusion}
In the framework of thermodynamic dislocation theory for nonuniform plastic deformation, we have developed the interface conditions for various interactions of dislocations with the grain boundary, e.g., dislocation accumulation, absorption, and transfer. Interfacial dissipation in terms of rates for the mean and jump of plastic slip has been proposed. An evolution equation for grain boundary misorientation has been developed. The proposed interface dissipation yields two thresholds for the dislocation absorption process. The dislocation absorption leads to asymmetry in the distribution of plastic slip and non-redundant dislocation density on two sides of the grain boundary. It turns out that due to the additional accumulated density of non-redundant dislocations, kinematic work hardening also develops during the process of dislocation absorption. The extent of misorientation and the length of the dislocation absorption process is determined not only by the interfacial dissipation, but also by the slip system of the grains.

\end{document}